# Active and passive crosslinking of cytoskeleton scaffolds tune the effects of cell inclusions on composite structure

Katarina Matic[1]*, Nimisha Krishnan[2]*, Eric Frank[2], Michael Arellano[1], Aditya Sriram[1], Moumita Das[3], Megan T Valentine[4], Michael J Rust[5], Rae M Robertson-Anderson[1†], Jennifer L. Ross[2†]

1. University of San Diego, Department of Physics and Biophysics
2. Syracuse University, Department of Physics
3. Rochester Institute of Technology, School of Physics and Astronomy
4. University of California, Santa Barbara, Department of Mechanical Engineering
5. University of Chicago, Department of Molecular Genetics and Cell Biology

*contributed equally to this work

†randerson@sandiego.edu, jlr@syr.edu

## Abstract

Incorporating cells within active biomaterial scaffolds is a promising strategy to develop forefront materials that can autonomously sense, respond, and alter the scaffold in response to environmental cues or internal cell circuitry. Using dynamic biocompatible scaffolds that can self-alter their properties via crosslinking and motor-driven force-generation opens even greater avenues for actuation and control. However, the design principles associated with engineering active scaffolds embedded with cells are not well established. To address this challenge, we design a dynamic scaffold material of bacteria cells embedded within a composite cytoskeletal network of actin and microtubules that can be passively or actively crosslinked by either biotin-streptavidin or multimeric kinesin motors. Using quantitative microscopy, we demonstrate the ability to embed cells of volume fractions 0.4 – 2% throughout the network without compromising the structural integrity of the network or inhibiting crosslinking or motor-driven dynamics. Our findings suggest that both passive and active crosslinking promote entrainment of cells within the network, while depletion interactions play a more important role in uncrosslinked networks. Moreover, we show that large-scale structures emerge with the addition of cell fractions as low as 0.4%, but these structures do not influence the microscale structural lengthscale of the materials. Our work highlights the potential of our composite biomaterial in designing autonomous materials controlled by cells, and provides a roadmap for effectively coupling cells to complex composite materials with an eye towards using cells as in situ factories to program material modifications.

# 1. Introduction

The future of materials engineering is to endow materials with adaptable, deformable, sensory, and responsive properties more akin to biological systems. In order to understand the design constraints on this class of materials, we seek to create and characterize prototype material systems that can leverage biological functions to achieve this design goal.

One particularly intriguing composite system for this goal is the cytoskeleton, comprising stiff microtubules, semiflexible actin filaments, and flexible intermediate filaments. It has been clear for decades that the cytoskeleton gives the cell shape, mechanical resilience, and adaptability as the individual components can organize and reorganize in space and time on the fly. More recently, in vitro reconstitution of composites of different cytoskeletal filaments, such as actin and microtubules, have revealed desirable emergent mechanical and structural properties that have not been able to be achieved with single-component networks (*1–7*). For example, passive actin-microtubule networks have been shown to exhibit increased mechanical resistance compared to actin networks as well as reduced local buckling and heterogeneity compared to microtubule networks (*1, 8*). Myosin-driven actin-microtubule composites have also been shown to exhibit more organized and tunable contractility compared to actomyosin networks without microtubules (*9, 10*). These studies and several others have now well characterized actin-microtubule composites (*1–5, 11–17*) including the effects of adding passive crosslinkers that alter the viscoelastic properties, and active crosslinking motors, including myosin and kinesin, to generate forces and restructure the composites (*6, 9, 10, 18, 19*). Moreover, studies have shown that these active elements can be externally triggered to change the composite organization, offering enhanced spatiotemporal control over activity and insights into how energy-consuming or catalytically-active systems couple to the mechanical systems (*9, 20, 21*).

The state-of-the art for introducing external stimuli in cytoskeletal networks is via light activation (*9, 20, 21*) which has allowed for triggered changes in activity and structure (*21, 22*), but ultimately, it would be desirable to couple these mechanochemical systems to an internal trigger that can be both generated and controlled within the material itself. A promising route to achieve such internal signaling is through the use of synthetic biology approaches to engineer bacteria capable of manufacturing and producing network modulating compounds (*23–26*). A first step toward the broad goal of using bacteria to trigger changes to cytoskeletal structure and mechanics, we need to understand how to design and formulate composite materials consisting of bacteria and cytoskeletal proteins to ensure that cells can be uniformly dispersed, and the surrounding network maintains its structural and dynamic properties.

Here, we characterize the effects of incorporating *E. coli* bacteria cells into interpenetrating networks of actin and microtubules. We find that cells at volume fractions of 0.4 – 2% are able to be well-integrated in the cytoskeleton scaffolds without loss of network integrity or significant alterations to mesh size. Moreover, we show that crosslinking microtubules, either with passive biotin-NeutrAvidin bonds or active tetrameric kinesin complexes, promotes entrainment of the cells by the network. This effect is evidenced by increased colocalization of cells and filaments as well as active dynamics of cells that mirror those of the network. Finally, we reveal that the presence of even the lowest cell fraction, leads to large-scale network remodeling, but this effect does not influence or undermine the network connectivity and structural uniformity on smaller lengthscales.

## 2. Materials and Methods

### 2.1 Cytoskeleton and Cells

*Cytoskeleton proteins*: We purchased lyophilized rabbit skeletal actin monomers (AKL99-C), rhodamine-labeled actin monomers (AR05-C), porcine brain tubulin dimers (T240), HiLyte647-labeled tubulin dimers (TL670M), and biotinylated tubulin dimers (T240) from Cytoskeleton, Inc. We reconstituted all proteins in PEM-100 [100 mM piperazine-N,N'-bis(ethane sulfonic acid) (pH 6.8), 2 mM $MgCl_2$, and 2 mM glycol ether diamine tetraacetic acid (EGTA)], and stored as single-use aliquots at -80°C.

*Crosslinkers*: To prepare passive crosslinkers, we preassembled complexes of NeutrAvidin, biotin, and biotinylated tubulin dimers at 1:2:2 ratio, as described previously (*27*). To prepare active crosslinkers, we purified biotinylated kinesin-401 expressed in Rosetta (DE3) pLysS competent E. coli cells (ThermoFisher), which we stored at -80°C in single-use aliquots, as described previously (*28*). Immediately prior to experiments, we prepared kinesin clusters by incubating kinesin-401 dimers with NeutrAvidin (ThermoFisher) at a 2:1 ratio in the presence of 4 µM DTT for 30 min at 4°C. In all experiments, crosslinkers (passive or active) are included in the sample at a final crosslinker to tubulin molar ratio of R = 0.04.

*Cells*: We used JM109 E. coli cells, which are a poorly motile strain with flagella (*29*). Bacteria carry the pGFP vector (Takara Biosciences) to allow it to express green fluorescent protein (GFP) in an inducible manner. Cell cultures were grown from glycerol stocks in LB at 30°C with 100 µg/mL ampicillin added to select for cells carrying pGFP. We monitored cell growth by measuring the optical density at 600 nm (OD) using a spectrophotometer. To induce GFP expression, we added 10 mM Isopropyl β-D-1-thiogalactopyranoside (IPTG) when the culture reached OD=0.04, and then we continued growth until reaching OD=0.1, after which we centrifuged 1 mL of the culture at 10,000g for 5 mins to pellet the cells. We then removed the supernatant and quantified the volume of the cell pellet, which we assumed to have the same cell density for all cultures. Before use in experiments, cells were inspected using transmitted and fluorescence imaging to ensure that they were all expressing GFP.

To achieve specific cell volume fractions, $\phi_c$ = 0.004, 0.008, 0.015, 0.023 in the networks, we reconstituted the cell pellets in varying volumes of LB, which we empirically determined from images. Specifically, to determine the necessary volume of LB to achieve the target $\phi_c$, we manually counted the number of cells $n_c$ across multiple images obtained by confocal microscopy of cell culture solution with the same initial cell volume (see Sections 2.2 and 2.3 for sample preparation and imaging details). We determined the volume represented by each image in which we counted cells as $V_I$ = 212 µm x 212 µm x 0.5 µm = $2.25\times10^4$ µm³ where 0.5 µm is the z-depth of each image. By analyzing the same images, we estimated the volume of a single cell to be $V_c \approx$ 1.4 µm³ (SI Fig. S1). We determined the volume fraction as $\phi_c = n_c V_c / V_I$ and used this expression to calibrate the relative dilutions. We stored resuspended cells at -20°C in single-use aliquots prior to use in composite material preparation.

## 2.2 Sample preparation

*Composite Preparation*: To prepare composites of cytoskeleton filaments and cells, we mixed actin monomers and tubulin dimers to final concentrations of 2.9 **µM** and 6.9 **µM**, with labeled:unlabeled subunit ratios of 1:20 and 1:10, respectively in PEM-100 supplemented with 4 mM GTP, 1 mM ATP, 5 µM paclitaxel, 4 µM phalloidin, 0.1% Tween20, and an oxygen scavenging system (45 µg mL$^{-1}$ glucose, 43 µg mL$^{-1}$ glucose oxidase, 7 µg mL$^{-1}$ catalase, 0.05% β-mercaptoethanol, and 5 µM Trolox). We then mixed in 2 µL of resuspended cells to achieve final cell volume fractions of $\phi_c = 0$, 0.004, 0.008, 0.015, 0.023. For $\phi_c = 0$, 2 µL of LB without cells was added. To polymerize the network, we incubated the sample at 30°C for 45 mins.

For networks with passive crosslinkers, we added biotin-NeutrAvidin complexes to the mixture prior to polymerization (*27*), which we carried out in the experimental sample chamber (see below). For networks with active crosslinkers (kinesin), because kinesin activity starts immediately upon adding to the network, we added kinesin clusters and 9 mM ATP following polymerization, which we carried out in a centrifuge tube, immediately prior to loading into the sample chamber and imaging. We also prepared composites with kinesin but without adding additional ATP. Because the networks have actin, which requires ATP to form, we always have some amount of ATP present, so these networks were also images immediately after inserting into the sample to reduce the amount of rearrangement that residual ATP might cause in the system.

For experiments, we introduced the sample by capillary flow into a chamber consisting of a glass coverslip and microscope slide separated by 500 µm by parafilm spacer and sealed with UV-curable glue. To passivate the chamber walls to prevent non-specific absorption of proteins or cells, we incubated the sample chamber with 150 mM BSA in PEM-100 for 10 minutes, used compressed air to force out the BSA solution and fully dried the chamber prior to inserting the sample.

### 2.3 Fluorescence Imaging

We performed experiments at both high and low magnification to determine structural properties of composites across a range of lengthscales. For high magnification imaging, we used a Nikon A1R laser scanning confocal microscope with a 60x 1.4 NA objective to collect stacks of 2D images at different z planes. We collected stacks of 81 2D images of size 512x512 pixel$^2$ (212 x 212 µm$^2$), each separated by a z-height of 0.5 µm, for a total stack height of 40 µm. We simultaneously recorded separate images for cells (GFP) (488 nm), actin (561 nm), and microtubules (647 nm), using 488 nm, 561 nm, and 640 nm laser lines and 520 nm, 593 nm and 670 nm emission filters. Images were acquired at 0.933 frames per second using galvanometer scanning with a pixel dwell time of 1.4 µs. For each composite type and cell concentration, we imaged three different samples and collected five image stacks in different x-y positions for each sample.

For low magnification imaging, we used a Nikon Ti-eclipse microscope with a Yokogawa CSU-W1 spinning disk confocal attachment, Plan Apo λ 10x objective, and Andor Zyla CMOS camera to collect 2D images. We collected time-series of images of size 2048x2048 pixel$^2$ (1331 x 1331 µm$^2$) with a 200 ms exposure time per frame, 30 second interval between frames, and a total time of 30 mins (61 frames). We simultaneously recorded separate images for cells (GFP) (488 nm), actin (561 nm), and microtubules (647 nm), using 488 nm, 561 nm, and 640 nm laser lines and 520 nm, 593 nm and 670 nm emission filters. Information on replicates and samples sizes can be found in in SI Table S1.

### 2.4 Quantitative Image Analysis

We performed All analyses described below were performed on each image of each stack for high-magnification data and the first frame of each low-magnification time-series. For analyses focused on structure and not dynamics, we focused the low-magnification structural analysis on the first frame to limit the extent to which the active crosslinkers (kinesin) used in these experiments have reorganized the network in the presence of residual ATP. This approach allowed us to approximately isolate the crosslinking role of kinesin from its active restructuring capability, to compare it to biotin-NeutrAvidin passive crosslinkers used in high-magnification experiments. For analysis on dynamics, we used the low-magnification images, as the large-scale restructuring was easier to capture for longer on the larger scale.

## 2.4.1. Spatial Image Autocorrelation

Spatial image autocorrelation (SIA) analysis was performed on both low and high magnification images using custom Python scripts to quantify the scaffold structure and cell distribution from the microscopy images. SIA produces an intensity autocorrelation $g(r)$, which is a measure of the correlation between two pixels separated by different radial distances $r$ (30). In general, $g(r)$ decays with increasing $r$, and can be evaluated to determine characteristic lengthscales at which pixel intensities become decorrelated, as described below. As previously described (10), we generate autocorrelation curves, by taking the fast Fourier transform of an image, multiplying it by its complex conjugate, applying an inverse Fourier transport and normalizing by the squared intensity:

$$g(r) = \frac{F^{-1}(|F(I(r))|^2)}{[I(r)]^2} \qquad (2)$$

Where $F$ represents the Fourier transform, $F^{-1}$ is the inverse Fourier transform, and $I(r)$ is the intensity of the image as a function of the distance, $r$. An average $g(r)$ function was calculated for each condition from the $g(r)$ curves computed for each image collected for that condition. To extract characteristic correlation length scales from the average $g(r)$ curves, we mask the $r = 0$ value $g(0)$, which is by definition 1, and fit the truncated data to a double exponential function

$$y = \Xi_s exp(-r/\xi_s) + \Xi_l exp(-r/\xi_l) \qquad (3)$$

where $\xi_s$ is the shorter characteristic exponential lengthscale with a weighting of $\Xi_s$, and $\xi_l$ is the longer characteristic exponential lengthscale with a weighting of $\Xi_l$. SIA analysis was performed on both the cell and network color channels. For some data, the best fit was a single exponential with a single characteristic length scale and amplitude.

## 2.4.2. Colocalization analysis

Colocalization analysis was used to examine the spatial colocalization of cells and cytoskeleton filaments in both low and high magnification images using a custom Python code as follows. The pixel intensities of each channel are rescaled as:

$$\tilde{I}(x,y) = \frac{(I(x,y) - I_{min}) - \langle (I(x,y) - I_{min}) \rangle}{I_{max} - I_{min}} \qquad (4)$$

where $I(x, y)$ is the intensity of a pixel at position $(x, y)$, $I_{min}$ and $I_{max}$ are the global minimum and maximum pixel value in the image, and the angled brackets denote the mean over all positions $(x, y)$. This process results in rescaled images for each channel: $\tilde{I}_c(x,y)$ (cells), $\tilde{I}_A(x,y)$ (actin), $\tilde{I}_M(x,y)$ (microtubules).

Colocalization between cells and filaments is assessed by multiplying the rescaled images of the corresponding channels to achieve colocalization images for actin and microtubules, $C_{c,A}(x,y) = \tilde{I}_c(x,y) * \tilde{I}_A(x,y)$ and $C_{c,M}(x,y) = \tilde{I}_c(x,y) * \tilde{I}_M(x,y)$

Colocalized images $C_{c,f}(x,y)$, where $f = A$ or $M$, are rescaled by their respective global minimum and maximum $C_{c,f,min}$ and $C_{c,f,max}$, similarly to the original images, via

$$\tilde{C}_{c,f}(x,y) = (C_{c,f}(x,y) - C_{c,f,min})/(C_{c,f,max} - C_{c,f,min}) \tag{5}$$

The resulting colocalization image has values that range between 0 and 1. To determine a single global colocalization parameter for each image, we compute the average across all pixel values $\langle \tilde{C}_{c,f}(x,y) \rangle$ where $\langle \tilde{C}_{c,f}(x,y) \rangle = 0$ and $\langle \tilde{C}_{c,f}(x,y) \rangle = 1$ indicate minimum colocalization and maximum colocalization, respectively.

### 2.4.3. Large scale structure characterization

To analyze large-scale network organization only observed in low-magnification videos, we measure the areas of the structures visible in the first frame. We perform this analysis by using the polygon selection tool in ImageJ, to manually outline the boundaries of the large structures and measure their areas. By normalizing the structure areas by the image area, we compute the fractional area.

### 2.4.4 Particle tracking

We quantify the mobility of the bacterial cells within the cytoskeletal networks using standard particle-tracking algorithms based on TrackPy, the Python implementation of colloidal particle tracking algorithms from Crocker and Grier (*31*). As previously described and implemented (*32–34*), we track the frame-to-frame displacements of all cells in $x$ and $y$ directions, from which we determine the corresponding mean-squared displacement as function of lag time $\tau$ for the ensemble of cells: $\langle (\Delta x(\tau))^2 \rangle$ and $\langle (\Delta y(\tau))^2 \rangle$. We compute the mean-squared displacement for each channel of each video as $MSD(\tau) = \frac{1}{2}[\langle(\Delta x(\tau))^2\rangle + \langle(\Delta y(\tau))^2\rangle]$

To determine the type and rate of motion, we fit each MSD to the power-law function $MSD(\tau) = K\tau^\alpha$ where $\alpha$ is the anomalous scaling exponent and $K$ is the generalized transport coefficient. For normal Brownian motion, $\alpha = 1$ and $K = 2D$ where $D$ is the diffusion coefficient. For ballistic motion, $\alpha = 2$ and $K = v$ where $v$ is the speed. Subdiffusive and superdiffusive motion is characterized by $\alpha < 1$ and $1 < \alpha < 2$, respectively.

### 2.4.5. Optical flow

To quantify the dynamics of both cells and filaments within active networks, we implemented the Farneback optical flow algorithm, using the function

"cv.calcOpticalFlowFarneback" from OpenCV (*35*). The output is a stack of 2D arrays of velocity vectors that represent the flow fields at different times and separated by a set frame interval. For each frame pair of the form $(i, i+n)$, where $i$ is a frame between the first and last frames in a video and $n$ is the number of frames separating the pair, the format of the output flow field is an array of shape $(H, W, 2)$, where $H$ and $W$ are the height and width of the image and the third dimension is a 2D velocity vector $\vec{u} = (u_x, u_y)$ where $u_x$ and $u_y$ are the $x$ and $y$ components of the velocity. To improve signal to noise, velocity vectors within each 10x10 square-pixel window comprising the image are then averaged together, to achieve a down-sampled array of $(H/10, W/10, 2)$. To calculate the mean velocity for each frame pair $i$, we average together all $u_x$ and $u_y$ values, resulting in a 2D average velocity vector $\langle \vec{u} \rangle = (\langle u_x \rangle, \langle u_y \rangle)$. We compute the average of $\langle \vec{u} \rangle$ across all frame pairs $i$, $\overline{\langle \vec{u} \rangle}$, and determine the average speed by computing the magnitude $\overline{\langle u \rangle} = \left[ \overline{\langle u_x \rangle}^2 + \overline{\langle u_y \rangle}^2 \right]^{1/2}$.

## 3. Results and discussion

We seek to investigate the structural and dynamic properties of mechanochemical composites of cytoskeleton networks embedded with bacteria cells that could ultimately regulate network organization and mechanics (Fig. 1). The cytoskeletal systems themselves are inherently complex composites comprising microtubule filaments assembled from tubulin dimers and actin filaments assembled from actin monomers. We also incorporate passive microtubule crosslinkers formed from neutravidin and biotinylated tubulin dimers, or active microtubule crosslinkers formed from multimeric kinesin motor clusters (Fig. 1A). Similar composite systems have been well characterized in previous works, as described in the introduction (*1*, *3*, *8–10*, *33*), offering roadmaps for tuning structure and mechanics for desired material properties. We designed the cytoskeleton network to have a mesh size of ~0.75 µm (details in SI Section S2) (*1*), which we chose to be comparable but slightly smaller than the size of the *E. coli* cells which are cylindrical objects of ~2.5 µm in length and ~0.85 µm in width (SI Fig. S1, SI Section 1). With this size matching, we expect that the bacterial cells should be incorporated into network architecture without too much disruption, but still sterically interact with the network (Fig. 1B). We characterize the networks using multi-color quantitative fluorescence microscopy using high resolution, high magnification (Fig. 1C) and lower magnification (Fig. 1D) imaging to characterize the organization of both the network and cells at different length scales. We use high-resolution optical sectioning enabled by confocal imaging and large-scale time-lapsed measurements to characterize the structure and dynamics of the networks in space and time.

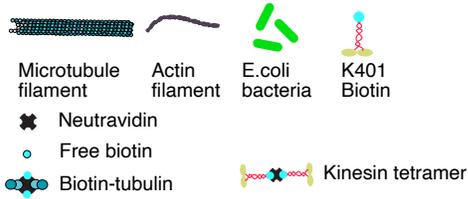
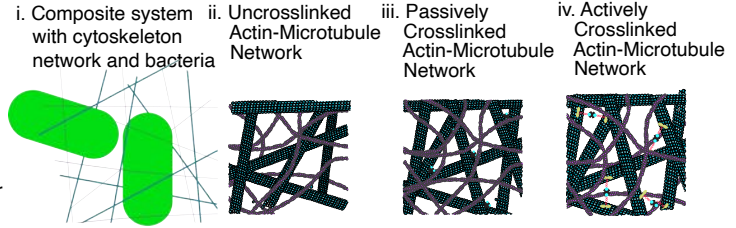
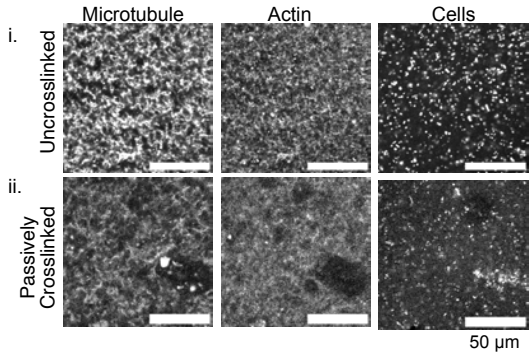
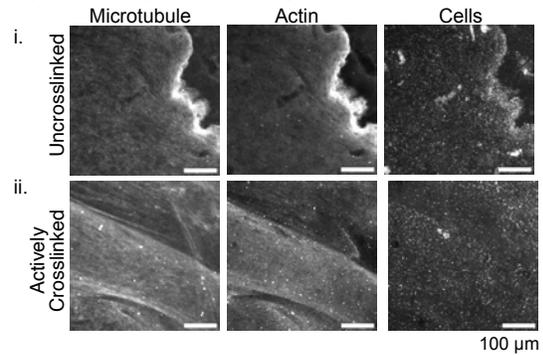
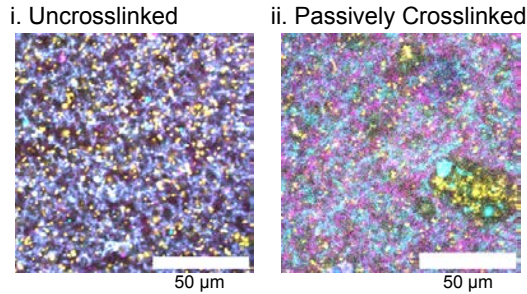
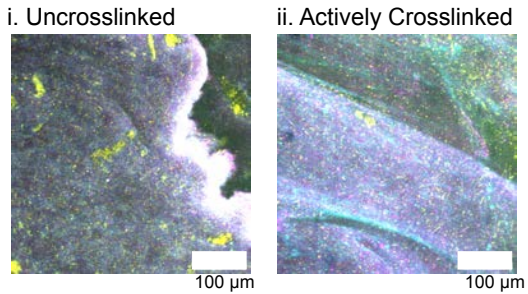

**Figure 1: Engineering composite materials of cytoskeletal filaments and bacteria cells.** (A) Cartoon diagram of the components of the composite including microtubules (cyan), actin filaments (magenta), and *E.coli* bacteria (green). Microtubules within the composites are either uncrosslinked, passively crosslinked using neutravidin to link biotinylated tubulin dimers, or actively crosslinked using kinesin tetramers. (B) Cartoons of composite networks. (i) To-scale schematic showing the relative sizes of the bacteria (green), microtubules (cyan) and acting filaments (magenta) and the mesh size of the filaments in composites. (ii-iv) Schematics of microtubule-actin networks (not to scale) with (ii) no crosslinkers, (iii) passive crosslinkers, and (iv) active crosslinkers. (C,D) Separate channels of images of composites at high (C) and low (D) magnification, without (i) and with (ii) passive (C) or active (D) crosslinkers, showing microtubules (left), actin (middle) and bacteria cells (right). (E,F) Merged color overlay of images from (C,D) showing microtubules (cyan), actin (magenta), and cells (yellow) for (i) uncrosslinked and (ii) either passively (E) or actively (F) crosslinked networks. Scale bars are 50 µm for high magnification images (C,E) and 100 µm for low magnification images (D,F). All the images shown are for $\phi_c$ = 0.015.

___

## 3.1 Crosslinking and high concentrations of bacteria cause network rearrangement in composites

To determine the effect of embedded cells on composite structure, we design composites with varying cell volume fractions $\phi_c = 0$, 0.004, 0.008, 0.015, and 0.023 (see Methods, Fig S1). We first analyze high magnification images to determine the impact of cells on the microscale structure of composites with and without passive crosslinkers. From qualitative visual inspection, we find that composites without crosslinkers show minimal impact of cells on composite structure for all cell densities (Fig. 2Ai, SI Fig. S2). However, the addition of passive crosslinkers leads to more clustering of cells and voids in the network (Fig. 2Aii, SI Fig. S3). To quantify the structural effects of cells and crosslinking, we perform spatial image autocorrelation (SIA), as described in Methods, to generate autocorrelation curves $g(r)$ which we evaluate to determine characteristic structural lengthscales of the composites. The different structural properties that crosslinking confers to all composite components is evident in the autocorrelation curves shown in Figure 2B for all components. Crosslinking generally increases the extent of spatial correlations at larger distances for both cells and filaments. As described in Methods, to quantify the structural correlation lengthscales, we fit each curve to a sum of two exponentials (Fig. 2B) to determine the short and long characteristic decay lengthscales, $\xi_s$ and $\xi_l$, respectively (Fig. 2C) and their respective amplitudes, $\Xi_s$ and $\Xi_l$ (Fig. 2D).

We find that actin and microtubules have similar short lengthscales of $\xi_s \approx 0.5$ - 2 µm which are insensitive to cell concentration and crosslinking (Fig. 2Ci,ii). This lengthscale is also comparable the network mesh size of ~0.75 µm. The larger length scales for actin and microtubules are also similar between uncrosslinked and crosslinked networks, but do display some dependence on cell volume fraction and filament type. Specifically, both actin and microtubules have $\xi_l \approx 1$ - 3 µm in the absence of cells, but the addition of even the lowest cell density increases $\xi_l$ for actin to ~6 µm (Fig. 2Cii). More modest dependence is seen for microtubules without a clear trend with cell density (Fig. 2Ci).

Turning to the structural properties of the cells, we find that cells also exhibit two characteristic length scales with the shorter being $\xi_s \approx 1$ µm for both crosslinked and uncrosslinked networks across all cell densities, similar to the filament networks (Fig. 2Ciii). This smaller lengthscale may be characteristic of the inherent size of the cells themselves, which are rounded cylinders of 2.5 µm and width of 0.8 µm in size (Fig. 2Ciii). In contrast to the filaments, the long lengthscales for the cells have a strong dependence on filament crosslinking. Specifically, for uncrosslinked networks, $\xi_l \approx 6$ µm for all cell densities, comparable to $\xi_l$ for actin. However, crosslinking increases $\xi_l$ to ~10-15 µm, substantially larger than any other lengthscales measured in the composites. We interpret $\xi_l$ for cells as characterizing the spacing between the bacteria cells. While SIA is unable to detect clusters of cells that we observe visually, due to the relatively low frequency of these events, we expect this clustering to result in larger average spacing between cells which are in different clusters.

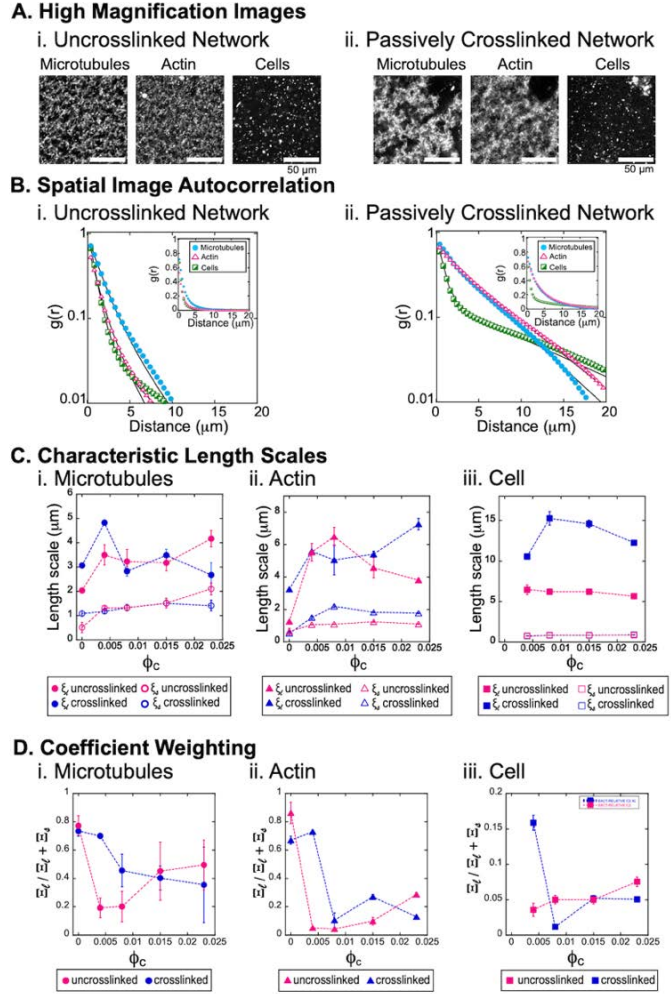

**Figure 2. Embedding cells in cytoskeleton networks cause structural changes at mesoscopic scales**. (A) Images of composite networks with bacteria at $\phi_c = 0.004$ for (i) uncrosslinked and (ii) passively crosslinked composites where each shows the microtubule (left), actin (center) and cell (right) channels. (B) Structural quantification is determined using spatial image autocorrelation to determine the autocorrelation function $g(r)$ for (i) uncrosslinked and (ii) passively crosslinked networks. For each, the plot is shown on a log-linear scale, with insets showing the same data plotted on a linear-linear scale. Lines show fits of the data to a sum of two exponentials (eq. 1). (C) The long, $\xi_l$ (filled symbols) and short $\xi_s$ (open symbols) characteristic length scales determined from the fits to g(r) and plotted as a function of cell volume fraction, $\phi_c$ for uncrosslinked (blue symbols) and crosslinked (magenta symbols) networks for (i) microtubules, (ii) actin, and (iii) bacteria cells. (D) The ratio of the long length scale coefficient (weight), $\Xi_l$, over the sum of the coefficients, $(\Xi_l + \Xi_s)$, determined from the fits to $g(r)$ and plotted as a function of $\phi_c$ for uncrosslinked (blue symbols) and crosslinked (magenta symbols) networks for (i) microtubules, (ii) actin, and (iii) bacteria cells. N values for all datasets can be found in SI Table S1 and error bars represent standard error.

_______________________________________________________________________________

To determine the relative significance of the two lengthscales for each condition, we evaluate the coefficient associated with the corresponding exponential term, $\Xi_l$ and $\Xi_s$ (see Methods). We can think of these coefficients as weights describing how important each length scale is to describing the network structure, which we quantify by computing the relative weight of the long length scale, $\Xi_l/(\Xi_l + \Xi_s)$ (Fig. 2D). This quantity can range from 0 to 1 for composites in which the short or long lengthscale respectively dominates the structure.

For the microtubules and actin, the long length scale has higher weighting when no cells are present (Fig. 2Di,ii), and the addition of even a small volume fraction of cells is enough to significantly reduce this weighting (Fig. 2Di,ii) (Fig. 2Di,ii, magenta). This result, along with the increase in $\xi_l$ upon addition of cells, suggests that cells may cause small scale bundling of filaments via depletion interactions between cells and filaments. Namely, bundling would increase $\xi_l$ by increasing the largescale spacing between bundled structures (i.e., more filaments per bundle result in larger distances between bundles). At the same time, there would be fewer individual fibers (bundles) contributing to the signal so the relative weighting is lower. We explore the role of depletion in composites further below.

Unlike for the filaments, in which the large lengthscale dominates the structure (i.e., $\Xi_l/(\Xi_l + \Xi_s)$ >0.5) at low cell densities (<0.01), the relative weighting of the large lengthscale for cells is <0.2 for all conditions (Fig. 2D,iii), demonstrating that the organization of the cells is dominated by the short length scales. This result suggests that cell clusters that contribute to the large lengthscales are few and far between, and the majority of cells are individually dispersed throughout the composite.

In all cases, the SIA analysis shows that, despite the images looking similar (SI Fig. S2,S3) even a very low concentration of bacteria cells in the network is sufficient to elicit quantitative structural effects at mesoscopic scales (i.e, several times the mesh size and cell size, $\xi_l$) while maintaining similar microscopic structure (i.e., $\xi_s$).

This scale-dependent impact of cells on composite structure, motivated us to examine composites at much larger lengthscales to determine if the structural effects of cells are amplified further at these scales. Inspecting images of composites without crosslinkers that span ~6x larger lengthscales, we observe that filaments and cells form large-scale patterns not evident at high-magnification, even at the lowest cell volume fractions (Fig. 3A, SI Fig. S4,S5). To examine the impact of crosslinking at these length scales we replace passive biotin-NeutrAvidin with a well-characterized multimeric kinesin construct that crosslinks microtubules and enables enzymatically-active remodeling of the network (Fig. 1A) (*36*). This also allowed us to observe dynamic restructuring that these active crosslinkers caused over time, which were not evident at high magnification (Fig. 3A, SI Fig. S4,S5).

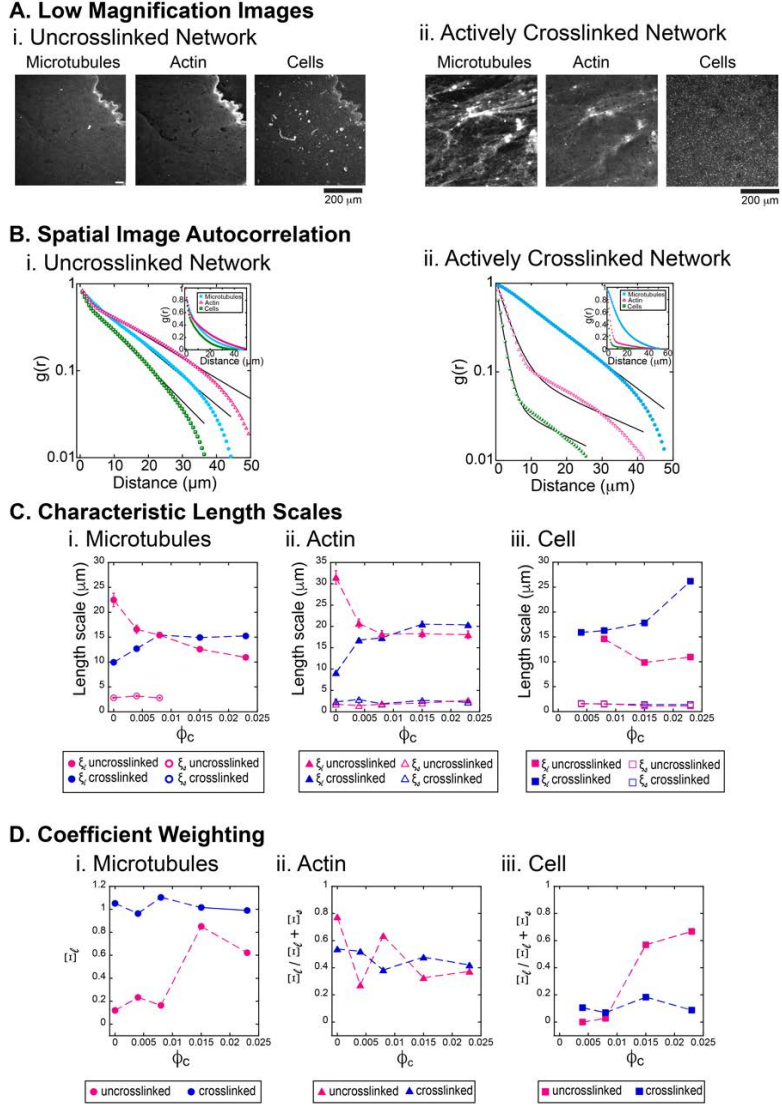

**Figure 3. Cells influence the large-scale network structure differently in uncrosslinked versus actively crosslinked composites.** (A) Images of composites with active crosslinkers and $\phi_c = 0.015$, showing the microtubule (left), actin (center) and cell (right) channels. (B) Autocorrelation curves $g(r)$ (symbols) and corresponding fits of the data to eq 1 (lines), shown on log-linear (left) and linear-linear (right scales). (C) The characteristic long and short length scales $\xi_l$ (filled symbols) and $\xi_s$ (open symbols), determined from the fits and plotted as function of $\phi_c$, are shown for the uncrosslinked (blue symbols) and crosslinked (magenta symbols) networks for (i) microtubules, (ii) actin, and (iii) bacteria cells. (D) Quantification of the coefficients $\Xi_l$ and $\Xi_s$ determined from the fits to $g(r)$ for uncrosslinked networks (blue symbols) and actively crosslinked networks (magenta symbols), plotted as the ratio of the long length scale coefficient, $\Xi_l$, over the sum of the coefficients, $(\Xi_l + \Xi_s)$. (i) For microtubules, most of the data was best fit to a single exponential, with a single coefficient $\Xi_l$, such that $\frac{\Xi_l}{\Xi_l + \Xi_s} = 1$. All data for actin (ii) and cells (iii) were best fit by a sum of two exponentials such that $0 < \frac{\Xi_l}{\Xi_l + \Xi_s} < 1$. N values for all dataset can be found in SI Table S1 and error bars represent standard error of the mean.

___

Using the same SIA analysis approach described above we examine both actively crosslinked networks and uncrosslinked networks (Fig. 3B). To facilitate comparison of our actively crosslinked composite results to the passively crosslinked cases, we restrict our analysis to the first frame of each time-series in which there could be active restructuring from residual ATP.

Similar to the high magnification data, we find that a sum of two exponentials fits most of the data well. The exception is microtubules in actively crosslinked networks (Fig. 3Ci, blue) and in uncrosslinked networks at higher cell concentrations (Fig. 3Ci, magenta). For all cases in which the data displays two lengthscales, the shorter lengthscales are $\xi_s \approx$ 1-3 µm for both filaments (Fig. 3Ci,ii) and $\xi_s \approx$ 1 µm for cells (Fig. 3Ciii), similar to those measured at high-magnification. We interpret this lengthscale as a measure of the size of the network mesh and an individual cell, respectively. Conversely, the larger lengthscales for both filaments and cells are substantially larger than their high magnification values, with values of $\xi_l \approx$ 10 - 30 µm (Fig. 3Ci,ii, solid). This effect may reflect the sensitivity of SIA to the finite size of the imaged field of view, with lower magnification imaging providing access to larger lengthscale structures and reduced measurement noise from more statistics. While high magnification images may be more sensitive to small-scale clustering or structures, captured by $\xi_l$, low magnification imaging can better capture large scale structures, also captured by $\xi_l$.

For both filament types, we find the uncrosslinked networks display the largest $\xi_l$ values in the absence of cells and this value decreases to an approximately constant value of $\xi_l \approx$ 15 µm as the cell density increases beyond $\phi_c \approx$ 0.005 (Fig. 3Cii, solid magenta). Interestingly, actively crosslinked networks are also sensitive to the cell volume fraction, but with an opposite trend. Without cells, the long length scale is significantly smaller than for the uncrosslinked composite, $\xi_l \approx$ 10 µm for both filament types, but increases to a plateau value similar to that of the uncrosslinked case. In the presence of cells at $\phi_c >$ 0.005, the effect of crosslinking becomes negligible for all cases. Without cells, we may expect the smaller $\xi_l$ for actively crosslinked networks to arise from bundling of filaments into local dense regions, whereas without crosslinkers, filaments can form large amorphous regions of entangled filaments that are locally homogeneous and $\xi_l$ may reflect the size of these regions. Adding cells to the networks can lead to depletion-driven bundling of uncrosslinked networks, which are freer to move and rearrange subject to entropic forces. The observation that the effect of cells on actin filaments is larger than for microtubules supports this conjecture as actin filaments are more flexible and can more readily rearrange in response to entropic forces. The same depletion forces could serve to have the opposite effect on crosslinked networks, driving small-scale bundles that form from crosslinkers to cluster and form larger bundles and structures, thereby increasing $\xi_l$.

Examining the large structural lengthscales for cells (Fig. 3Ciii) we find that, similar to high magnification, $\xi_l$ values for crosslinked composites are generally larger than for uncrosslinked composites, but this difference is reduced compared to the high magnification case and is only significant at cell volume fractions above ~0.01. Unlike the passively crosslinked networks, however, the longer length scale of the cells embedded in actively crosslinked networks appears to increase with increasing cell concentration. This trend is consistent with the relative insensitivity of $\xi_l$ on cell density above $\phi_c > 0.005$ for both filaments. As more cells are added to the networks, they have little impact on the network structure, but instead lead to growing clusters of cells.

As in figure 2D, we also quantify the relative coefficients of the fit terms to determine the relative importance of the long and short characteristic lengthscales to the composite structure (Fig. 3D). For the microtubule conditions in which the data was better fit by a single exponential with a lengthscale comparable to the other measured $\xi_l$ values, the relative coefficient is $\frac{\Xi_l}{(\Xi_l+\Xi_s)} = 1$ ($\Xi_s = 0$) (Fig. 3Di), implying that the structure is nearly completely dominated by large scale organization. Conversely, for microtubules in uncrosslinked networks at low cell densities (above $\phi_c < 0.01$), the contribution from the large length scale is quite low, at $\Xi_l/(\Xi_l + \Xi_s) \approx 0.2$, suggesting low formation of largescale structures.

For actin, we find that for both crosslinked and uncrosslinked networks, the relative weighting for the long length scale was ~0.5, indicating that the short and long lengthscales contribute equally to the structure of the actin network in the composites (Fig. 3Dii). Moreover, $\Xi_l/(\Xi_l + \Xi_s)$ is relatively insensitive to cell concentration. These results are quite distinct from the high magnification trends for $\Xi_l$ in which the weighting transitions from high values (>0.5) to low values (<0.5) with the addition of cells (Fig. 2D); and suggest that the largescale structure of actin is relatively decoupled from the restructuring of microtubules and cells.

The weighting analysis for cells show that at low cell concentrations, the organization of cells is dominated by the short structural length scale, since the long length scale weighting is ~0.2 (Fig. 3Diii, blue), suggesting that the cells are mostly dispersed single cells embedded in the network. Indeed, this appears to be the case from inspection (SI Fig. S5). For the uncrosslinked network, the long length scale begins to dominate the structure for $\phi_c > 0.01$ (Fig. 3Diii, magenta), in opposition to the high magnification case in which the weighting is reduced at higher $\phi_c$. These results suggest that clumping or other organization of cells becomes important at higher cell volume fractions, but because of their large size, their contribution to the high magnification structure is reduced while it is increased in low magnification images. Importantly, this increased weighting of the large lengthscale is not observed in crosslinked composites which maintain $\Xi_l/(\Xi_l + \Xi_s) \approx 0.1$ for all cell densities. This result is important because it shows that the cells only stay well separated when the composite is crosslinked, implying that crosslinking of the composite is likely necessary to keep the cells embedded and homogeneously separated within the network. This will be important for future studies planning to use the bacteria to control the network connectivity, organization, and mechanics.

Overall, the composite cytoskeleton of microtubules and actin combined with bacteria cells were generally able to create a network that could embed and separate bacteria cells over 2% of the volume fraction. Surprisingly, the bacteria cells had some effects on the network both at small and large scales, even at very low cell volume fractions, 0.004 and 0.008. For both the organization of the network and the cells, crosslinking appeared to help maintain the organization as more cells were added, although structural changes to the filaments were still observed above 1% (v:v) of cells included.

Taken together, our results demonstrate that we are able to successfully generate composite scaffolds of microtubules and actin that when combined with bacteria cells demonstrate good mixing and maintain relatively uniform distributions up to cell volume fractions over 2%. Upon careful inspection, we identified modest effects in network structure that were lengthscale dependent and evident even at very low cell volume fractions (<1%). However, we do not observe any obvious aggregation, demixing, or other phase separation behaviors that would undermine the mechanical resiliency or performance of the material.

The lengthscale dependent remodeling we observe may be explained by entropically-driven depletion interactions, in which aggregation of larger polymers or filaments can be induced via crowding by smaller colloidal particles or polymers (*34*, *37–39*). Indeed, the effects of crowding on bundling of cytoskeleton networks has been demonstrated previously (*39–41*). However, many of these studies use much higher volume fractions of inclusions. Prior works that included similarly small volumes of micron-scale inert beads for mechanical measurements have not reported restructuring effects of micron-sized inclusions into similar cytoskeletal composite networks (*1*, *3*, *8*, *32*, *33*). However, there are reports of local depletion of polymer filaments near surfaces, particularly for semiflexible and rigid filaments, such as actin and microtubules, due to 'self-depletion' effects, which arise from surface-induced steric constraints (*42*). The length scale over which these effects appear correlates to the filament length, with depletion zones up to ~35 μm reported in measurements of actin near planar glass surfaces (*42*). Such depletion effects have also been observed through microrheology measurements, particularly when the filament length is similar to the diameter of the colloidal particle (*43*). In this limit, a local softening of the network rheology is observed due to the local reduction in polymer concentration near the particle surface. However, the lengths of our filaments are ~5-10 μm, several times larger than the cell length, so it is unlikely this excluded volume effect plays an important role. As we conjecture above, it is more likely that the differences we observe in the long length scale arise from changes in filament bundling, which could further exacerbate the steric constraints at the cell surface. Further, bundling could arise from the activity of the kinesins in the presence of the residual amount of ATP or from changes in network mobility due to crosslinking, both of which could create local heterogeneities.

Alternatively, it is possible that the bacteria are inducing these changes due to some other mechanism, because they are not just inert colloids. They have non-motile flagella, which are filaments projecting from their surfaces that allow them to adhere to surfaces and swim when motile. Further, the cells themselves have surface patterns of charge and hydrophobic groups, which could result in non-steric filament interactions (*44*).

### 3.2 Cytoskeleton network crosslinkers increase interactions with bacterial cells

Our SIA analysis described above demonstrated that adding bacterial cells at increasing concentrations were able to modify the network while remaining relatively well separated both at small and large length scales. If the cells are becoming entrapped in the network, we might also expect to see increased steric interactions and increased colocalization of cytoskeletal components and cells with increasing cell density. Conversely, if cells were causing depletion driven de-mixing and clustering then we would expect to see no increase in cell-filament colocalization upon increasing cell concentration.

To quantify the interaction between the filaments and the cells, we perform quantitative colocalization analysis of the images, comparing the cell fluorescence channel with that of each type of cytoskeletal filament, as described in Methods Section 2.4.2 (Fig. 4). Specifically, we calculate a unique colocalization metric for each filament type, $C_A$ and $C_M$, in each condition and at each magnification. This metric can range between 0 and 1 for complete separation or maximal observed colocalization between filaments and cells, respectively.

Examining these colocalization metrics (Fig. 4A,B), we found qualitative and quantitative differences for crosslinked composites compared to uncrosslinked composites at both magnifications (Fig. 4B,C). At high magnification, the colocalization of bacteria with microtubules or actin was low and insensitive to cell concentration for uncrosslinked networks (Fig. 4Ci, magenta). Crosslinking caused a significant increase in this colocalization and adding more cells caused a further increase in the colocalization (Fig. 4Ci, blue). These results suggest that depletion effects may be more significant in uncrosslinked networks, whereas crosslinking promotes entrainment of cells within networks. This physical picture is consistent with the fact that the large lengthscale for cells is significantly larger in crosslinked networks than in uncrosslinked networks (Fig. 2Ciii) as they are able to spread into the space occupied by the network.

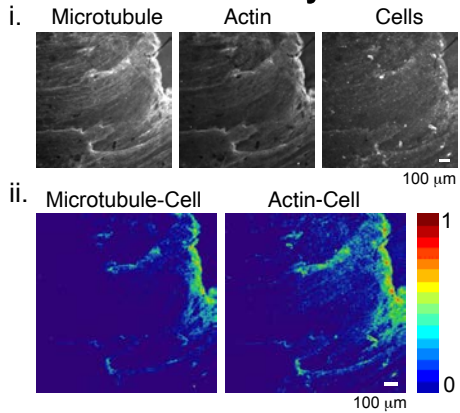
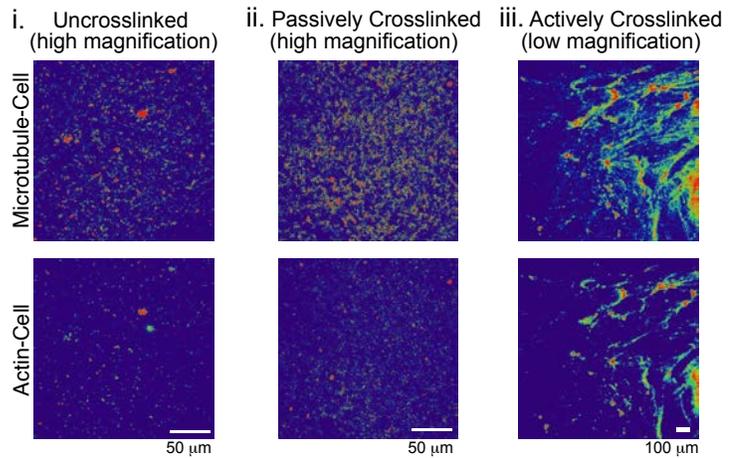
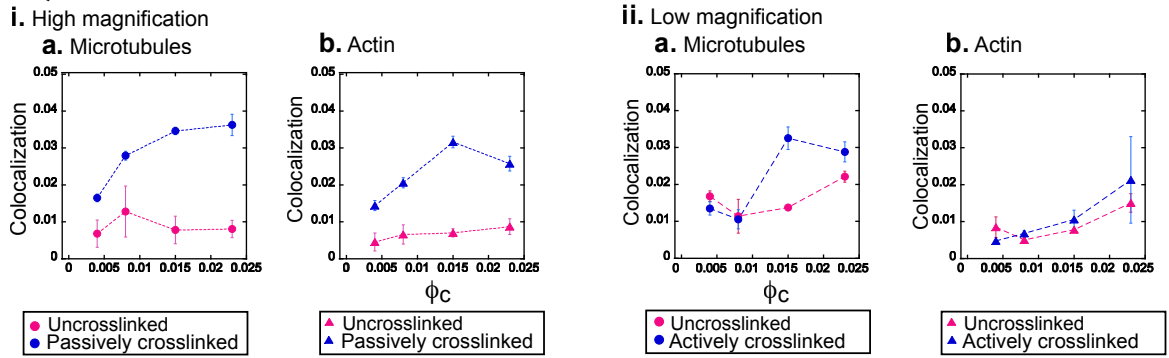

**Figure 4. Colocalization of cytoskeletal filaments with bacteria cells is enhanced by crosslinking.** (A) Colocalization analysis method example. (i) Examples images at low magnification for microtubules (left), actin (middle), and cells (right) without any crosslinker at $\phi_c = 0.023$. Scale bar is 100 µm and applies to all images. (ii) Colocalization images for microtubules interacting with cells (left) and actin interacting with cells (right) using the same images as in (i). The color look-up table shows the range of values of co-localization metrics $C_A$ and $C_M$ from 0 (blue) to 1 (red). Scale bar is 100 µm and applies to all images. (B) Example colocalization images for different network compositions and magnifications: (i) uncrosslinked networks at high magnification, (ii) passively crosslinked networks at high magnification, and (iii) actively crosslinked networks at low magnification. The top and bottom rows shows microtubule-cell colocalization $C_M$ and actin-cell colocalization $C_A$, respectively. (C) Quantification of colocalization metrics for microtubules (a) and actin (b), averaged over multiple images and chambers for various cell volume fractions imaged at (i) high and (ii) low magnification. Each plot shows data for uncrosslinked (blue filled circles) and crosslinked (pink filled circles) networks. Crosslinkers are either passive (i) or active (ii). N values for all datasets can be found in SI Table S1 and error bars represent standard error of the mean.

______________________________________________________________________

Similar to our high magnification results, at low magnification we observe enhanced colocalization between microtubules and cells in crosslinked networks compared to uncrosslinked networks, but only at higher cell concentrations ($\phi_c > 0.01$) (Fig. 4Ciia). This higher cell concentration for the onset of enhanced colocalization makes sense considering the larger lengthscales over which colocalization must occur to be captured at lower magnification. Conversely, actin and cells appear to be only modestly colocalized at low magnification for both network types all cell concentrations (Fig. 4Ciib). This result indicates that cells are interacting more strongly with microtubules than actin, which aligns with the increased colocalization with crosslinking we observe, as it is the microtubules in the network that are crosslinked, and it is this crosslinking that likely 'cages' and entrains the cells. (Fig. 4Biii,Cii).

### 3.3 Entrained cells cause large-scale structured domains to form in cytoskeleton networks

The low magnification images shown in figures 1, 3, 4 reveal largescale feature that appear to have some structure, rather than the isotropic structure we observe at high magnification (SI Fig. S4,S5). Microtubule and actin filaments are polar filaments with a high aspect ratio, microns long and nanometers wide. Due to this inherently high aspect ratio, cytoskeletal filaments can act like liquid crystal mesogens that can align at high concentration (*36*, *45–53*). Indeed, actin and microtubules are easy to bundle with crowding agents through depletion interactions and in addition to specific crosslinkers (*37*, *39*, *40*, *54*, *55*). As the bacteria concentration increases, we might expect that the crowding due to the presence of these large, cylindrical colloid-like particles could cause increased local density, as we describe above, as well as alignment of the filaments. To investigate the local alignment of microtubules and actin in composites, we visually examine the low-magnification images to identify regions of local gradients and alignment.

For this analysis, we focus on low magnification images to maximize our observed field of view. At high magnification, it is difficult to determine the boundaries of any larger regions, which often appear to be larger than the observed field, and all filaments within the volume appear to be largely isotropically entangled, even in the case of passively crosslinked networks (SI Fig. S2,S3). However, at low magnification, structures on a larger scale can be observed for both uncrosslinked and crosslinked networks and appear to be amplified by increasing amounts of bacteria (Fig. 5A). Importantly, in the absence of bacteria ($\phi_c$ =0), we observe no large-scale structures. The uncrosslinked networks are homogeneous and isotropic, and the addition of active crosslinkers creates local punctate structures that are homogeneously distributed across the field of view (Fig. 5A, left).

When bacteria are added, we observe the formation of large-scale structures for both uncrosslinked and actively crosslinked networks, even at the lowest cell density ($\phi_c$ = 0.004) (Fig. 5A, violet box). Active crosslinking appears to modestly enhance this effect (Fig. 5A). To more quantitatively assess the formation of large-scale structures, we quantified the fractional areas of the structured regions within the networks (see Methods), which span the entire imaging area in some cases (Fig. 5Bi). Any areas that appeared to be homogeneous and unstructured were not included in the fractional area assessment. As observed qualitatively, the addition of cells at any concentration can create structured domains with and without crosslinkers (Fig. 5Bii). This effect is more robust across cell densities when crosslinkers are added, with many of the networks showing that >80% of the imaging area is in a structured region (Fig. 5Biii). This result is consistent with the increased colocalization we observe in crosslinked networks (Fig. 4). Cells that are better integrated into the network may have a more pronounced effect on the structure of the filament-rich domains.

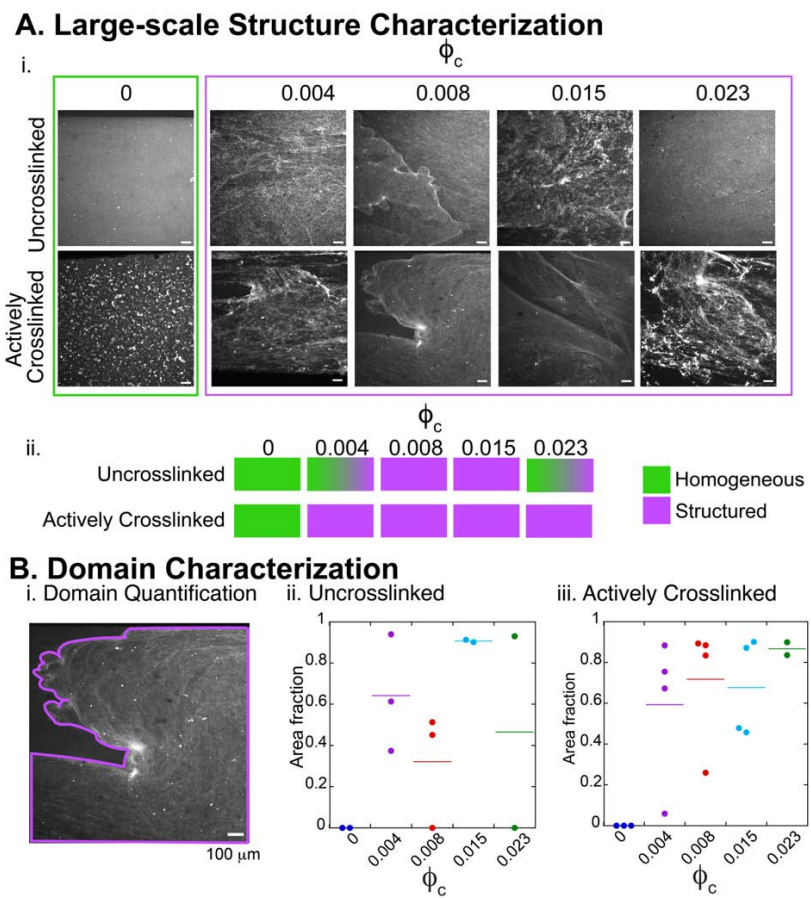

**Figure 5. Characterization of large-scale structures in cell-cytoskeleton composites.** (A) (i) Example images of microtubules in uncrosslinked (top row) and actively crosslinked (bottom row) composite networks with increasing cell volume fraction (listed above each image). Images are characterized as homogeneous (green border) or structured (violet border). (ii) This classification is displayed as color blocks that represent the organization in each condition, with the results of individual samples for each condition displayed within the corresponding block. (B) Characterization of the area of structured domains. (i) Example image of microtubules in an actively crosslinked network at $\phi_c = 0.008$, with the structured region outlined in violet. (ii,iii) Quantification of the areas of structured domains for uncrosslinked (ii) and actively crosslinked (iii) networks as a function of $\phi_c$. The circles are areas measured for each image and the horizontal lines denote the average. N values for all datasets can be found in SI Table S1 and error bars denote standard error.

___________________________________________________________________

## 3.4 Bacteria cells are entrained in actively restructuring cytoskeleton networks

The long-term goal for this research is to embed programmable bacteria within active cytoskeletal networks to enable autonomous and/or triggered responses of the biomaterial. This goal requires the cells to be well-mixed with the cytoskeletal networks, which we demonstrate above. Additionally, the bacteria must also remain within the network even during active motor-driven remodeling. To further explore the dynamic properties of the material and assess the stability of mixing, we examine the effects of kinesin-driven activity, again using the low-magnification imaging to capture large fields of view. As described above, our analysis shown in Figures 3-5 focused on the first frame of the videos we acquired for the actively crosslinked networks to attempt to isolate the role of crosslinking by the kinesin without considering enzymatic activity. Here, we analyze the whole videos to characterize the time-varying composite structure and the mobility of the networks and the cells within them.

We observe large scale changes of both the network and the bacteria cells that are entrained in the network, which we characterize with a temporal overlay where each frame over 30 minutes is a different color. As seen in the representative colormap (Fig. 6A), the networks move unidirectionally and the cells clearly move with the networks. To quantify the motion of the actin, microtubules, and cells in the networks we use optical flow to generate velocity vector fields and compute average speeds for each component, as described in the Methods (*56*). We find that both cytoskeletal components as well as the cells move with the same velocity, which is roughly constant over increasing cell volume fractions at ~20 nm/s (Fig. 6B). This result demonstrates that cells are indeed entrained in the network and can couple to the active motion of the filaments.

Optical flow assumes ballistic motion between frame intervals, an assumption that may not be accurate for the cells that likely have contributions from thermal fluctuations in addition to being entrained with the actively moving filaments. To more accurately characterize the motion of the cells, we use particle-tracking algorithms (see Methods) to track the trajectories of the cells, which are bright punctate objects ideal for particle-tracking (Fig. 6Ci). For each condition, we compute the mean squared displacement (MSD) of the ensemble of tracked cells as a function of lag time $\tau$ (Fig. 6Cii). As described in Methods, we fit each MSD to a power-law function $MSD = K\tau^\alpha$, where $K$ is the generalized transport coefficient and $\alpha$ is the anomalous exponent. For normal Brownian motion, $\alpha = 1$ and $K = 2D$ where $D$ is the diffusion coefficient. For ballistic motion, $\alpha = 2$ and $K = v$ where $v$ is the speed. Superdiffusive motion is characterized by $1 < \alpha < 2$, respectively.

We find that cells exhibit superdiffusive dynamics across all cell densities, similar to previous reports of colloid dynamics in active cytoskeleton composites (*19*, *20*). However, at the highest cell fraction, the scaling exponent drops from $\alpha \approx 1.4$ to ~1.1, which is close to the exponent expected for purely diffusive behavior (Fig. 6Ciii). This effect may indicate that as the cell volume fraction becomes too high, many of the cells are excluded from the network, rather than being entrained, so the dynamics are largely from diffusive dynamics of the cells that are decoupled from the active network. Similarly, the generalized mobility constant, $K$, also depends on the cell volume fraction, with lower cell densities having a lower $K$ value than the highest cell volume fraction (Fig. 6Civ). This may seem counterintuitive and opposite from what is observed in al flow, but the units of $K$ depend on $\alpha$. The shift in the values of $K$ are indicative that there is a change from more ballistic to more diffusive behavior, where the ideal units for the diffusion coefficient are µm²/s, and the ideal units for ballistic motion are the same as velocity, µm/s. Moreover, at lower cell densities, we measure $K \approx 0.02$ µm²/s$^\alpha$ which equates to 20 µm/s for ballistic motion ($\alpha = 1$), consistent with our optical flow results (Fig 6B).

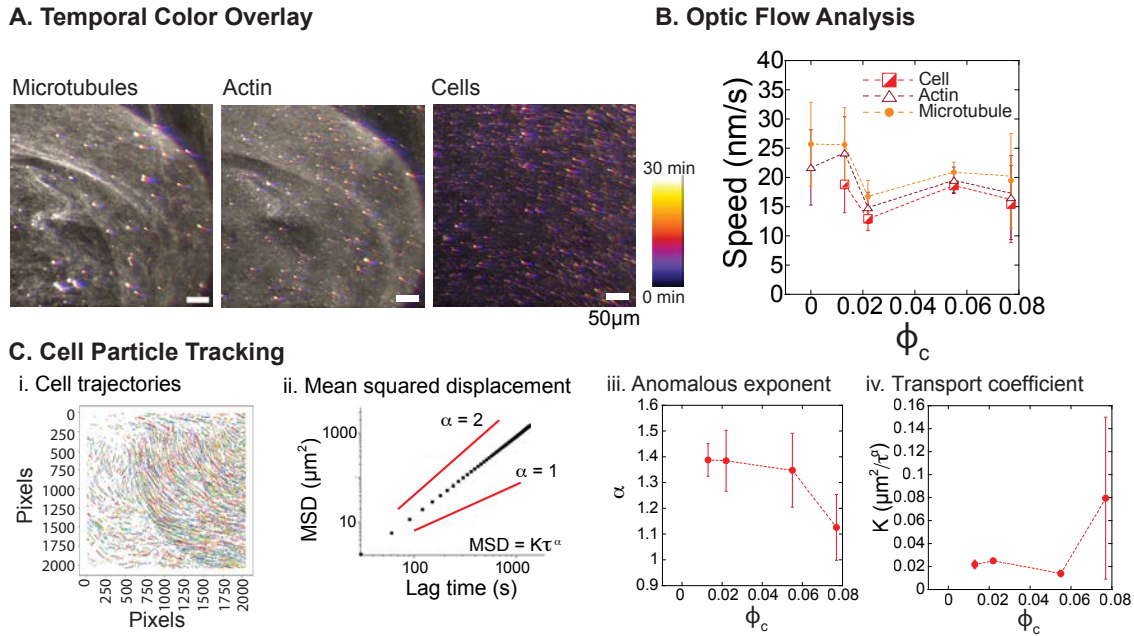

**Figure 6. Cells couple to the active dynamics of the cytoskeletal composites**. (A) Example of composite motion over time using color overlay showing microtubules (left), actin (middle), and cells (right) at $\phi_c = 0.015$. Different points in time (frames) are denoted by different colors according to the color scale shown that ranges from 0 min (black) to 30 min (white). Scale bar is 200 µm and applies to both images, which are cropped from the upper-right quarter of the original image. (B) Images are analyzed using optical flow to compute average speed as function of $\phi_c$ for microtubules (orange filled circles), actin (open red triangles), and cells (half-filled red squares) in a composite with active crosslinkers. (C) Bacteria cell motion is characterized by particle tracking. (i) Example trajectories of bacteria and (ii) mean-squared displacement MSD versus lag time $\tau$ for the movie shown in (A). (iii) Anomalous exponent $\alpha$ from fits of MSDs to the equation shown in (ii) plotted as a function of cell volume fraction for active composites. (iv) The generalized transport coefficient $K$ from fits of MSDs to the equation shown in (ii) plotted as a function of cell volume fraction for active composites. N values for all dataset can be found in SI Table S1 and error bars denote standard error.

___

## 4. Conclusions

A promising route for creating novel materials with programmable active properties is to embed them with active agents that can alter the surrounding material in response to internal circuitry and/or external cues. With an eye towards this design goal, our study takes an important first step by creating a composite material comprising a biopolymer scaffold embedded with bacteria cells. The integration of living bacteria cells into similar scaffolds can enable local control of the chemomechanical responses of the material by the cells acting as sensing and response centers. The results demonstrate our ability to engineer a composite network of microtubules and actin that can entrain bacteria cells while retaining structural integrity. We show that the characteristic structural lengthscales and the extent to which cells and filaments colocalize can be tuned by the addition of passive and active microtubule crosslinkers. We conjecture that this tuning is possible due to the varying impacts of depletion interactions and altered filament mobility in the different composites. Further, we show that this integration of cells within the networks is preserved during large-scale dynamic rearrangements driven by active kinesin crosslinkers, with the cell mobility tracking with that of the filaments for cell volume fractions up to 2%.

A particularly surprising result of our study is that even very low cell volume fractions, as low as 0.4%, can cause structural changes to the network which are amplified at larger lengthscales. Importantly, these rearrangements do not appear to affect the ability of the cells to be embedded or entrained in the networks, or the mesh size of the network. While crowding and depletion-driven restructuring of polymer networks by inclusions is a well-known mechanism for bundling of polymers, reminiscent of our results, these effects are typically observed at much higher volume fractions of crowders. To conjecture the origin of this unique behavior we observe, we note that even non-motile, non-growing bacteria cells, as we have here, are not simply inert spherical colloids. Their anisotropic shape and inhomogeneous surface properties may play a role in causing the lengthscale-dependent re-organization of the networks (*29, 57–60*). Our future work will focus on delineating the roles of steric colloidal interactions and the unique biochemistry of bacteria cells to the results we observe. Regardless of the mechanism, our study provides a blueprint for effectively coupling cells to complex composite materials, laying the foundation for the use of cells as in situ factories that can trigger programmable structural and mechanical changes of materials.

**Data Availability Statement:** Data for this article are available on Zenodo at https://doi.org/10.5281/zenodo.14552697

**Conflicts of Interest:** Authors declare no conflicts of interest.

**Author Contributions:**
Performed experiments: KM, NK, EF
Analyzed data: KM, NK, MA, AS, JLR, RMRA
Interpreted data: KM, NK, JLR, RMRA, MJR, MTV
Supervised research: RMRA, JLR
Provided guidance: MJR, MTV, MD
Wrote paper: JLR, KM, NK, RMRA
Edited the paper: MJR, MTV, MD, AS

**Acknowledgements:** The authors acknowledge Sharon Ferguson and Joseph Provost for providing the JM109 bacteria with the pGFP vector to make the cells fluorescent for imaging. MTV was supported by NSF DMREF-2118497; MD was supported by NSF DMREF-211-8449; MJR was supported by NSF DMREF-211-8424; KM, MA, AS, and RMRA were supported by NSF DMREF-2119663; NK, EF, and JLR were supported by NSF DMREF-2118403. KM and RMRA were supported by Arnold and Mabel Beckman Foundation Beckman Scholars Program.

**References:**

1. S. N. Ricketts, J. L. Ross, R. M. Robertson-Anderson, Co-Entangled Actin-Microtubule Composites Exhibit Tunable Stiffness and Power-Law Stress Relaxation. *Biophysical Journal* **115**, 1055–1067 (2018).

2. M. L. Francis, S. N. Ricketts, L. Farhadi, M. J. Rust, M. Das, J. L. Ross, R. M. Robertson-Anderson, Non-monotonic dependence of stiffness on actin crosslinking in cytoskeleton composites. *Soft Matter* **15**, 9056–9065 (2019).

3. S. N. Ricketts, M. L. Francis, L. Farhadi, M. J. Rust, M. Das, J. L. Ross, R. M. Robertson-Anderson, Varying crosslinking motifs drive the mesoscale mechanics of actin-microtubule composites. *Sci Rep* **9**, 12831 (2019).

4. S. N. Ricketts, P. Khanal, M. J. Rust, M. Das, J. L. Ross, R. M. Robertson-Anderson, Triggering Cation-Induced Contraction of Cytoskeleton Networks via Microfluidics. *Front. Phys.* **8**, 596699 (2020).

5. L. Farhadi, S. N. Ricketts, M. J. Rust, M. Das, R. M. Robertson-Anderson, J. L. Ross, Actin and microtubule crosslinkers tune mobility and control co-localization in a composite cytoskeletal network. *Soft Matter*, 10.1039.C9SM02400J (2020).

6. J. Berezney, B. L. Goode, S. Fraden, Z. Dogic, Extensile to contractile transition in active microtubule–actin composites generates layered asters with programmable lifetimes. *Proc. Natl. Acad. Sci. U.S.A.* **119**, e2115895119 (2022).

7. H. Wu, Y. Shen, S. Sivagurunathan, M. S. Weber, S. A. Adam, J. H. Shin, J. J. Fredberg, O. Medalia, R. Goldman, D. A. Weitz, Vimentin intermediate filaments and filamentous actin form unexpected interpenetrating networks that redefine the cell cortex. *Proc. Natl. Acad. Sci. U.S.A.* **119**, e2115217119 (2022).

8. S. J. Anderson, C. Matsuda, J. Garamella, K. R. Peddireddy, R. M. Robertson-Anderson, R. McGorty, Filament Rigidity Vies with Mesh Size in Determining Anomalous Diffusion in Cytoskeleton. *Biomacromolecules* **20**, 4380–4388 (2019).

9. G. Lee, G. Leech, M. J. Rust, M. Das, R. J. McGorty, J. L. Ross, R. M. Robertson-Anderson, Myosin-driven actin-microtubule networks exhibit self-organized contractile dynamics. *Sci. Adv.* **7**, eabe4334 (2021).

10. G. Lee, G. Leech, P. Lwin, J. Michel, C. Currie, M. J. Rust, J. L. Ross, R. J. McGorty, M. Das, R. M. Robertson-Anderson, Active cytoskeletal composites display emergent tunable contractility and restructuring. *Soft Matter* **17**, 10765–10776 (2021).

11. V. Pelletier, N. Gal, P. Fournier, M. L. Kilfoil, Microrheology of microtubule solutions and actin-microtubule composite networks. *Phys Rev Lett* **102**, 188303 (2009).

12. M. Dogterom, G. H. Koenderink, Actin–microtubule crosstalk in cell biology. *Nat Rev Mol Cell Biol* **20**, 38–54 (2019).

13. J. L. Henty-Ridilla, A. Rankova, J. A. Eskin, K. Kenny, B. L. Goode, Accelerated actin filament polymerization from microtubule plus ends. *Science* **352**, 1004–1009 (2016).

14. J. L. Henty-Ridilla, M. A. Juanes, B. L. Goode, Profilin Directly Promotes Microtubule Growth through Residues Mutated in Amyotrophic Lateral Sclerosis. *Current Biology* **27**, 3535-3543.e4 (2017).

15. J. Sabo, M. Dujava Zdimalova, P. G. Slater, V. Dostal, S. Herynek, L. Libusova, L. A. Lowery, M. Braun, Z. Lansky, CKAP5 enables formation of persistent actin bundles templated by dynamically instable microtubules. *Current Biology* **34**, 260-272.e7 (2024).

16. A. Elie, E. Prezel, C. Guérin, E. Denarier, S. Ramirez-Rios, L. Serre, A. Andrieux, A. Fourest-Lieuvin, L. Blanchoin, I. Arnal, Tau co-organizes dynamic microtubule and actin networks. *Sci Rep* **5**, 9964 (2015).

17. M. P. López, F. Huber, I. Grigoriev, M. O. Steinmetz, A. Akhmanova, G. H. Koenderink, M. Dogterom, Actin–microtubule coordination at growing microtubule ends. *Nature Communications* **5** (2014).

18. L. Farhadi, C. Fermino Do Rosario, E. P. Debold, A. Baskaran, J. L. Ross, Active Self-Organization of Actin-Microtubule Composite Self-Propelled Rods. *Frontiers in Physics* **6** (2018).

19. J. Y. Sheung, D. H. Achiriloaie, C. Currie, K. Peddireddy, A. Xie, J. Simon-Parker, G. Lee, M. J. Rust, M. Das, J. L. Ross, R. M. Robertson-Anderson, Motor-Driven Restructuring of Cytoskeleton Composites Leads to Tunable Time-Varying Elasticity. *ACS Macro Lett.* **10**, 1151–1158 (2021).

20. G. Lee, M. J. Rust, M. Das, R. J. McGorty, J. L. Ross, R. M. Robertson-Anderson, Myosin-driven actin-microtubule networks exhibit self-organized contractile dynamics. *Science Advances* **in press** (2020).

21. T. D. Ross, H. J. Lee, Z. Qu, R. A. Banks, R. Phillips, M. Thomson, Controlling organization and forces in active matter through optically defined boundaries. *Nature* **572**, 224–229 (2019).

22. M. Schuppler, F. C. Keber, M. Kröger, A. R. Bausch, Boundaries steer the contraction of active gels. *Nat Commun* **7**, 13120 (2016).

23. T.-C. Tang, B. An, Y. Huang, S. Vasikaran, Y. Wang, X. Jiang, T. K. Lu, C. Zhong, Materials design by synthetic biology. *Nat Rev Mater* **6**, 332–350 (2020).

24. A. Cubillos-Ruiz, T. Guo, A. Sokolovska, P. F. Miller, J. J. Collins, T. K. Lu, J. M. Lora, Engineering living therapeutics with synthetic biology. *Nat Rev Drug Discov* **20**, 941–960 (2021).

25. A. P. Liu, E. A. Appel, P. D. Ashby, B. M. Baker, E. Franco, L. Gu, K. Haynes, N. S. Joshi, A. M. Kloxin, P. H. J. Kouwer, J. Mittal, L. Morsut, V. Noireaux, S. Parekh, R. Schulman, S. K. Y. Tang, M. T. Valentine, S. L. Vega, W. Weber, N. Stephanopoulos, O. Chaudhuri, The living interface between synthetic biology and biomaterial design. *Nat. Mater.* **21**, 390–397 (2022).

26. Q. Wang, Z. Hu, Z. Li, T. Liu, G. Bian, Exploring the Application and Prospects of Synthetic Biology in Engineered Living Materials. *Advanced Materials*, 2305828 (2023).

27. S. N. Ricketts, M. L. Francis, L. Farhadi, M. J. Rust, M. Das, J. L. Ross, R. M. Robertson-Anderson, Varying crosslinking motifs drive the mesoscale mechanics of actin-microtubule composites. *Sci Rep* **9**, 12831 (2019).

28. R. J. McGorty, C. J. Currie, J. Michel, M. Sasanpour, C. Gunter, K. A. Lindsay, M. J. Rust, P. Katira, M. Das, J. L. Ross, R. M. Robertson-Anderson, Kinesin and myosin motors compete to drive rich multiphase dynamics in programmable cytoskeletal composites. *PNAS Nexus* **2**, pgad245 (2023).

29. M. M. Abdulkadieva, E. V. Sysolyatina, E. V. Vasilieva, A. I. Gusarov, P. A. Domnin, D. A. Slonova, Y. M. Stanishevskiy, M. M. Vasiliev, O. F. Petrov, S. A. Ermolaeva, Strain specific motility patterns and surface adhesion of virulent and probiotic Escherichia coli. *Sci Rep* **12**, 614 (2022).

30. C. Robertson, Theory and practical recommendations for autocorrelation-based image correlation spectroscopy. *J. Biomed. Opt* **17**, 080801 (2012).

31. J. C. Crocker, D. G. Grier, Methods of Digital Video Microscopy for Colloidal Studies. *Journal of Colloid and Interface Science* **179**, 298–310 (1996).

32. J. Y. Sheung, J. Garamella, S. K. Kahl, B. Y. Lee, R. J. McGorty, R. M. Robertson-Anderson, Motor-driven advection competes with crowding to drive spatiotemporally heterogeneous transport in cytoskeleton composites. *Front. Phys.* **10**, 1055441 (2022).

33. S. J. Anderson, J. Garamella, S. Adalbert, R. J. McGorty, R. M. Robertson-Anderson, Subtle changes in crosslinking drive diverse anomalous transport characteristics in actin-microtubule networks. *Soft Matter* **17**, 4375–4385 (2021).

34. R. M. Robertson-Anderson, *Biopolymer Networks: Design, Dynamics and Discovery* (Institute of Physics Publishing, Bristol, 1st ed., 2023)*Biophysical Society-IOP Series*.

35. G. Farnebäck, "Two-Frame Motion Estimation Based on Polynomial Expansion" in *Image Analysis*, J. Bigun, T. Gustavsson, Eds. (Springer Berlin Heidelberg, Berlin, Heidelberg, 2003; http://link.springer.com/10.1007/3-540-45103-X_50)vol. 2749 of *Lecture Notes in Computer Science*, pp. 363–370.

36. T. Sanchez, D. T. N. Chen, S. J. DeCamp, M. Heymann, Z. Dogic, Spontaneous motion in hierarchically assembled active matter. *Nature* **491**, 431–4 (2012).

37. R. Fitzpatrick, D. Michieletto, K. R. Peddireddy, C. Hauer, C. Kyrillos, B. J. Gurmessa, R. M. Robertson-Anderson, Synergistic Interactions Between DNA and Actin Trigger Emergent Viscoelastic Behavior. *Phys Rev Lett* **121**, 257801 (2018).

38. S. M. Gorczyca, C. D. Chapman, R. M. Robertson-Anderson, Universal scaling of crowding-induced DNA mobility is coupled with topology-dependent molecular compaction and elongation. *Soft Matter* **11**, 7762–8 (2015).

39. J. Clarke, L. Melcher, A. D. Crowell, F. Cavanna, J. R. Houser, K. Graham, A. M. Green, J. C. Stachowiak, T. M. Truskett, D. J. Milliron, A. M. Rosales, M. Das, J. Alvarado, Morphological control of bundled actin networks subject to fixed-mass depletion. *The Journal of Chemical Physics* **161**, 074905 (2024).

40. P. Chandrakar, J. Berezney, B. Lemma, B. Hishamunda, A. Berry, K.-T. Wu, R. Subramanian, J. Chung, D. Needleman, J. Gelles, Z. Dogic, Engineering stability, longevity, and miscibility of microtubule-based active fluids. *Soft Matter* **18**, 1825–1835 (2022).

41. R. Sakamoto, M. P. Murrell, F-actin architecture determines the conversion of chemical energy into mechanical work. *Nat Commun* **15**, 3444 (2024).

42. C. I. Fisher, S. C. Kuo, Filament rigidity causes F-actin depletion from nonbinding surfaces. *Proc. Natl. Acad. Sci. U.S.A.* **106**, 133–138 (2009).

43. H. Lee, J. M. Ferrer, F. Nakamura, M. J. Lang, R. D. Kamm, Passive and active microrheology for cross-linked F-actin networks in vitro. *Acta Biomaterialia* **6**, 1207–1218 (2010).

44. N. A. K. Bharadwaj, J. G. Kang, M. C. Hatzell, K. S. Schweizer, P. V. Braun, R. H. Ewoldt, Integration of colloids into a semi-flexible network of fibrin. *Soft Matter* **13**, 1430–1443 (2017).

45. T. Sanchez, D. Welch, D. Nicastro, Z. Dogic, Cilia-like beating of active microtubule bundles. *Science* **333**, 456–9 (2011).

46. A. J. M. Wollman, C. Sanchez-Cano, H. M. J. Carstairs, R. A. Cross, A. J. Turberfield, Transport and self-organization across different length scales powered by motor proteins and programmed by DNA. *Nature Nanotechnology* **9**, 44–47 (2014).

47. B. Edozie, S. Sahu, M. Pitta, A. Englert, C. F. do Rosario, J. L. Ross, Self-organization of spindle-like microtubule structures. *Soft Matter* **15**, 4797–4807 (2019).

48. J. Brugués, D. Needleman, Physical basis of spindle self-organization. *Proc Natl Acad Sci U S A* **111**, 18496–500 (2014).

49. J. Roostalu, J. Rickman, C. Thomas, F. Nédélec, T. Surrey, Determinants of Polar versus Nematic Organization in Networks of Dynamic Microtubules and Mitotic Motors. *Cell* **175**, 796-808.e14 (2018).

50. J. Viamontes, P. W. Oakes, J. X. Tang, Isotropic to Nematic Liquid Crystalline Phase Transition of F-Actin Varies from Continuous to First Order. *Physical Review Letters* **97** (2006).

51. K. L. Weirich, S. Banerjee, K. Dasbiswas, T. A. Witten, S. Vaikuntanathan, M. L. Gardel, Liquid behavior of cross-linked actin bundles. *Proceedings of the National Academy of Sciences* **114**, 2131–2136 (2017).

52. K. L. Weirich, K. Dasbiswas, T. A. Witten, S. Vaikuntanathan, M. L. Gardel, Self-organizing motors divide active liquid droplets. *Proc. Natl. Acad. Sci. U.S.A.* **116**, 11125–11130 (2019).

53. R. Zhang, N. Kumar, J. L. Ross, M. L. Gardel, J. J. de Pablo, Interplay of structure, elasticity, and dynamics in actin-based nematic materials. *Proc Natl Acad Sci USA* **115**, E124–E133 (2018).

54. J. L. Ross, D. K. Fygenson, Mobility of Taxol in Microtubule Bundles. *Biophysical Journal* **84**, 3959–3967 (2003).

55. D. J. Needleman, M. A. Ojeda-Lopez, U. Raviv, K. Ewert, H. P. Miller, L. Wilson, C. R. Safinya, Radial compression of microtubules and the mechanism of action of taxol and associated proteins. *Biophys. J.* **89**, 3410–3423 (2005).

56. D. Lee, The optic flow field: the foundation of vision. *Phil. Trans. R. Soc. Lond. B* **290**, 169–179 (1980).

57. M. M. Santore, Interplay of physico-chemical and mechanical bacteria-surface interactions with transport processes controls early biofilm growth: A review. *Advances in Colloid and Interface Science* **304**, 102665 (2022).

58. P. Prachaiyo, L. A. Mclandsborough, A Microscopic Method to Visualize Escherichia coli Interaction with Beef Muscle. *Journal of Food Protection* **63**, 427–433 (2000).

59. J. Won, J.-W. Kim, S. Kang, H. Choi, Transport and Adhesion of Escherichia coli JM109 in Soil Aquifer Treatment (SAT): One-Dimensional Column Study. *Environ Monit Assess* **129**, 9–18 (2007).

60. J. Li, L. A. McLandsborough, The effects of the surface charge and hydrophobicity of Escherichia coli on its adhesion to beef muscle. *International Journal of Food Microbiology* **53**, 185–193 (1999).

# Active and passive crosslinking of cytoskeleton scaffolds tune the effects of cell inclusions on composite structure

Katarina Matic[1]*, Nimisha Krishnan[2]*, Eric Frank[2], Michael Arellano[1], Aditya Sriram[1], Moumita Das[3], Megan T Valentine[4], Michael J Rust[5], Rae M Robertson-Anderson[1]†, Jennifer L. Ross[2]†

1. University of San Diego, Department of Physics and Biophysics
2. Syracuse University, Department of Physics
3. Rochester Institute of Technology, School of Physics and Astronomy
4. University of California, Santa Barbara, Department of Mechanical Engineering
5. University of Chicago, Department of Molecular Genetics and Cell Biology

*contributed equally to this work

†randerson@sandiego.edu, jlr@syr.edu

**Supplementary Information**

**S1. Cell volume fraction calculation**

To establish a predictive relationship between initial cell concentration of our cell cultures, and the cell volume fraction when we diluted fixes amounts, we performed two empirical measurements. First, we measured the average size of the bacteria cells using high magnification z-stack images using the scanning confocal microscope system (SI Fig S1A). The length and width of the bacteria was determined through analysis of the intensity profiles along the major and minor axes of the cylindrical crosssection at the z-slice closest to the midplane of the cell (SI Fig S1A). Using this method, we quantified that the length was 2.5 µm and width was 0.85 µm. The shape of the cell was assumed to be a spherically capped cylinder with two end caps that were idealized as perfect half-spheres. This gave a cell volume of 1.4 µm³.

Next, we used high magnification z-stack images taken from the scanning confocal system to estimate the number of cells $n_c$ within the imaging volume of $V_I = 212$ µm x 212 µm x 0.5 µm = $2.25 \times 10^4$ µm³ where 0.5 µm is the z-depth of each image. Specifically, we manually counted cells within a collapsed z-stack where the different frames are displayed as different colors (Fig S1Bi). For each dilution, 4 z-stacks of images were analyzed. We determined the volume fraction as $\phi_c = n_c V_c / V_I$ and used this expression to calibrate the relative dilutions. Specifically, we plot $\phi_c$ vs cell dilution and fit the data, which expected scale linearly, to a line, the slope of which gives the calibration factor (SI Fig. S1).

## A. Determination of Cell Volume
### i. Length
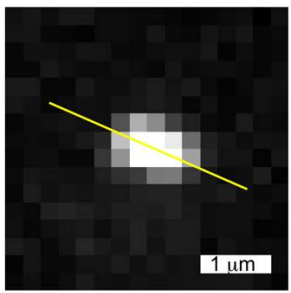
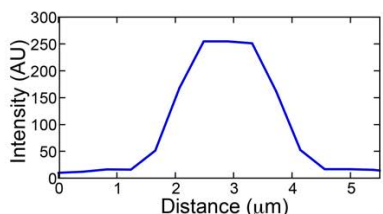

### ii. Width
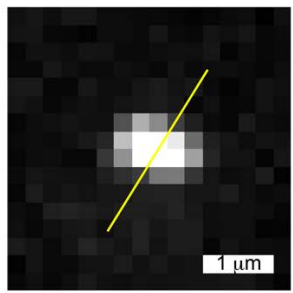
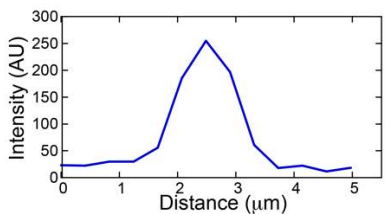

## B. Quantification of Cell Volume Fraction
### i. Projection of z-stack
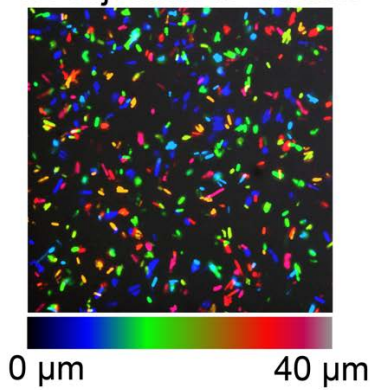

### ii. Cell volume fraction
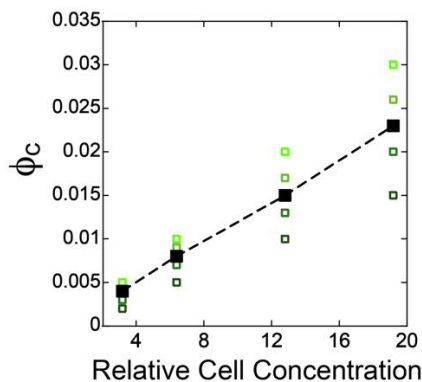

**Figure S1. Determination of cell volume and cell volume fraction.** (A) Cell volume was determined from direct measurements of images of cells embedded in networks. (i) Example of method to determine the cell length by drawing a line through the long axis and plotting the intensity profile. (ii) Example method to determine the cell width by drawing a line through the short axis and plotting the intensity profile. Scale bars are 5 µm; N=10 cells for each measurement. (B) Determination of the cell volume fraction. (i) Confocal z-stacks of networks with cells embedded in them collapsed with a color intensity profile with the spectrum look-up table. (ii) Cell volume fraction was determined from the number of cells counted in a given sample volume and converted to the total cell volume within the sample volume. Four different chambers were measured for each cell concentration used in experiments.

___

## S2. Mesh size calculation

Using these equations:

$$\zeta_C = (\zeta_A^{-3} + \zeta_M^{-3})^{-1/3} \qquad (S1)$$

$$\zeta_A = \frac{0.3}{c_A^{1/2}}, \qquad (S2)$$

$$\zeta_M = \frac{0.89}{c_M^{1/2}}, \qquad (S3)$$

we calculated the expected mesh size for the actin-microtubule composite network, $\zeta_C$, from the equation for the mesh size of the actin network, $\zeta_A$, and microtubule network, $\zeta_M$, where c$_A$ represents the concentration of actin in mg/mL, and c$_M$ represents the concentration of tubulin in mg/mL. We used a tubulin concentration of 0.76 mg/ml and an actin concentration of 0.12 mg/ml, yielding a composite mesh size of 0.75 µm.

## S3. Example images of microtubules, actin, and cells in networks for increasing cell volume fractions

### High magnification uncrosslinked network

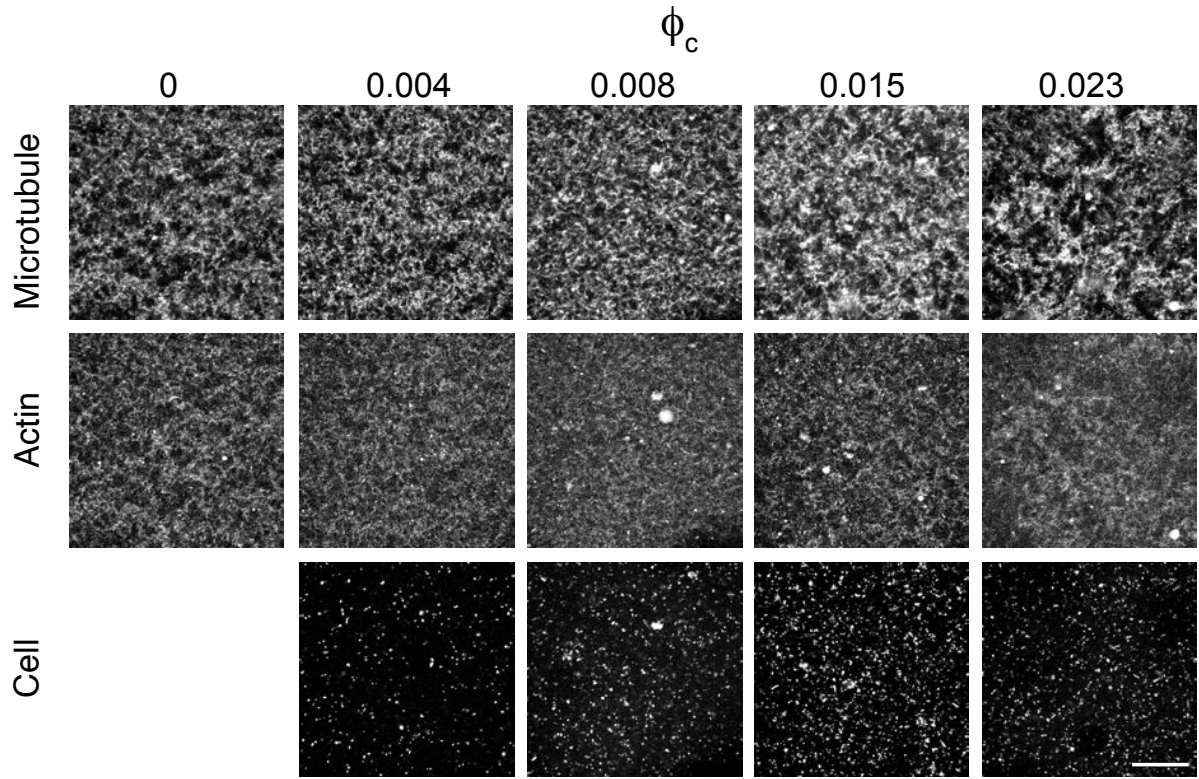

Fig S2. Representative images of **uncrosslinked** composite cytoskeleton and bacteria networks imaged at **high** magnification showing the (A) microtubule portion of the network, (B) the actin portion of the network, and the (C) cell portion of the network each with increasing cell volume fractions. Scale bar shown in one panel represents 50 μm and is the same for all panels.

## High magnification crosslinked network

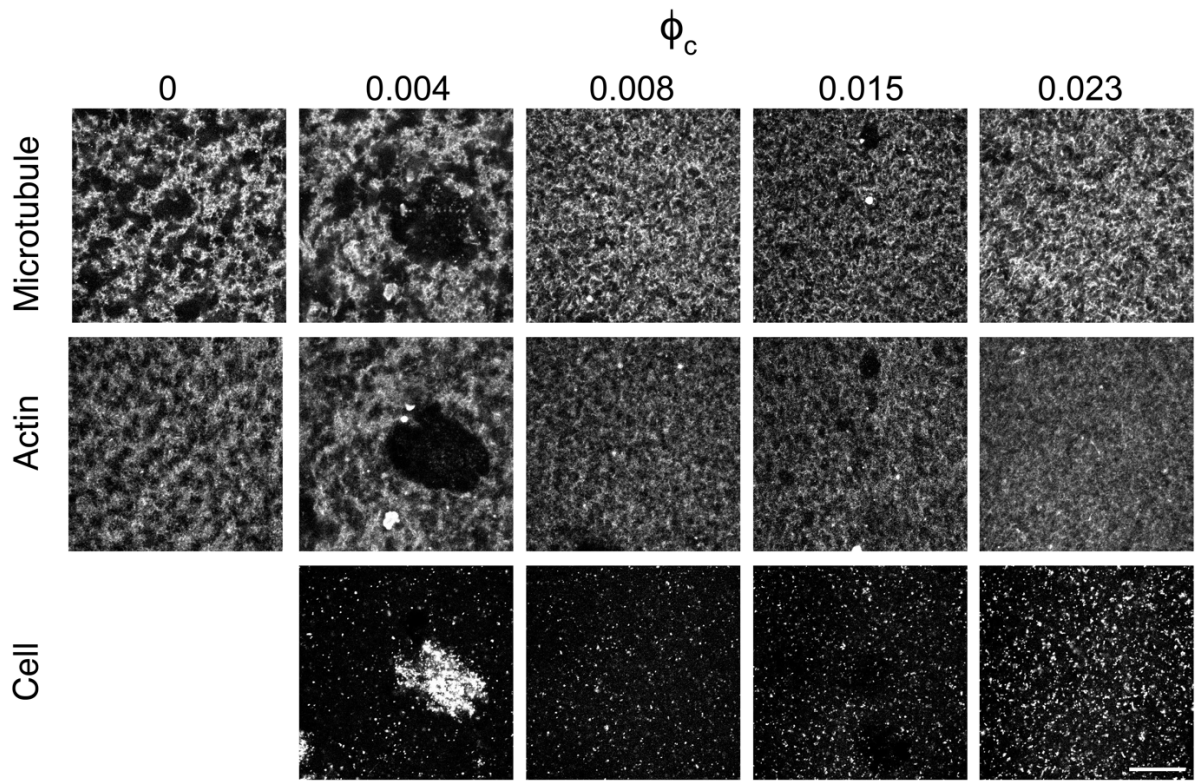

Fig S3. Representative images of **crosslinked** composite cytoskeleton and bacteria networks imaged at **high** magnification showing the (A) microtubule portion of the network, (B) the actin portion of the network, and the (C) cell portion of the network each with increasing cell volume fractions. Scale bar shown in one panel represents 50 μm and is the same for all panels.

## Low magnification uncrosslinked network

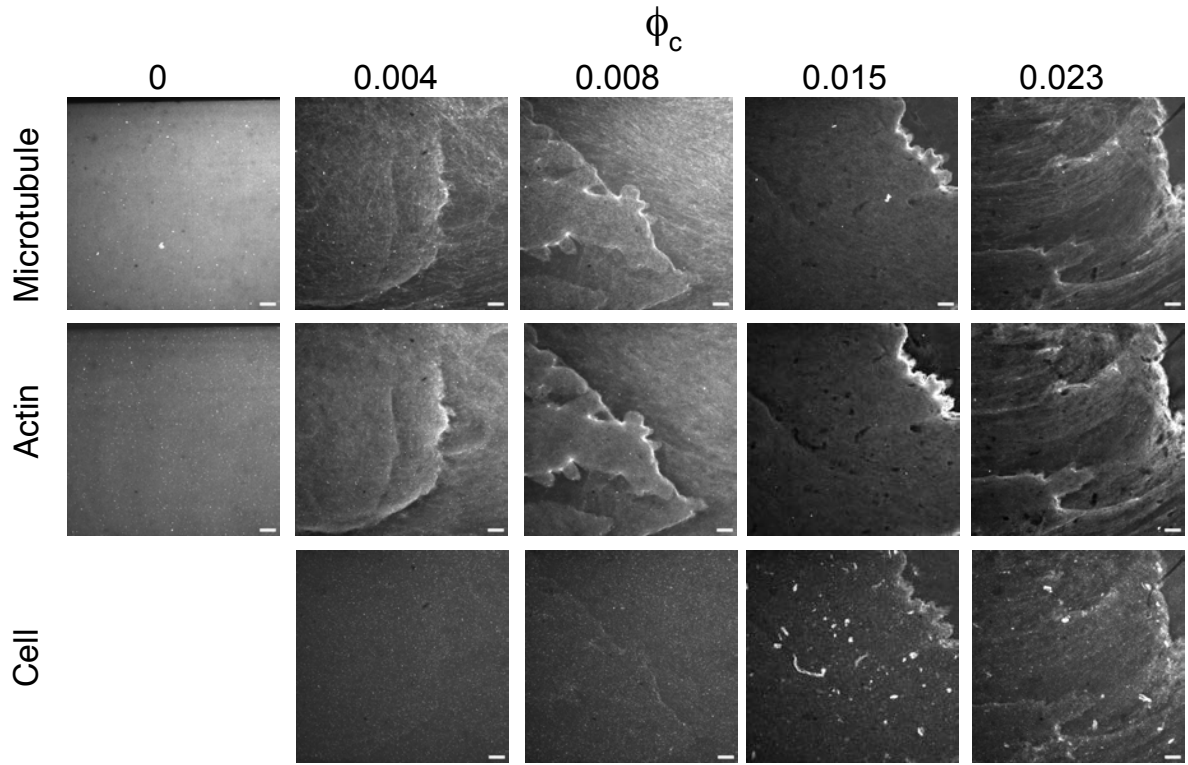

Fig S4. Representative images of **uncrosslinked** composite cytoskeleton and bacteria networks imaged at **low** magnification showing the (A) microtubule portion of the network, (B) the actin portion of the network, and the (C) cell portion of the network each with increasing cell volume fractions. Scale bars are the same for all images and represent 50 µm in length.

## Low magnification crosslinked network

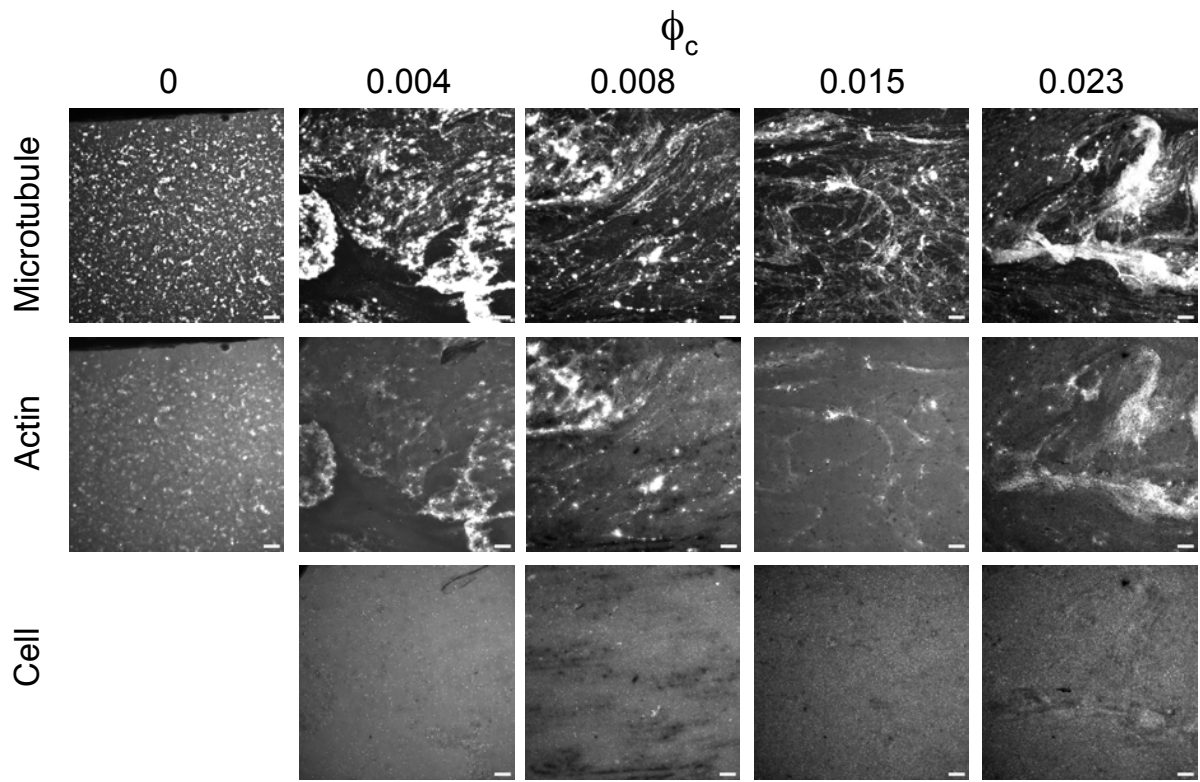

Fig S5. Representative images of **crosslinked** composite cytoskeleton and bacteria networks imaged at **low** magnification showing the (A) microtubule portion of the network, (B) the actin portion of the network, and the (C) cell portion of the network each with increasing cell volume fractions. Scale bars are the same for all images and represent 50 µm in length.

## S4. Number of samples for each condition

**Supplemental Table S1**

| Network type / magnification | Samples per $\phi_c =$ 0 / 0.004 / 0.008 / 0.015 / 0.023 |
|---|---|
| uncrosslinked / high magnification (60x) | 10 / 10 / 10 / 10 / 10 |
| passively crosslinked / high magnification (60x) | 10 / 10 / 10 / 10 / 10 |
| uncrosslinked / low magnification (10x) | 2 / 2 / 2 / 2 / 2 |
| actively crosslinked / low magnification (10x) | 3 / 3 / 3 / 3 / 3 |